\PassOptionsToPackage{svgnames}{xcolor}
\documentclass[12pt,reqno]{article}

\usepackage[numbers,sort&compress]{natbib}
\usepackage{amsthm, amsmath, amsfonts, amssymb, amscd, mathtools, youngtab, euscript, mathrsfs, verbatim, enumerate, multicol, multirow, bbding, color, babel, esint, geometry, tikz, tikz-cd, tikz-3dplot, array, enumitem, hyperref, thm-restate, thmtools, datetime, graphicx, tensor, braket, slashed, standalone, pgfplots, ytableau, subfigure, wrapfig, dsfont, setspace, wasysym, pifont, float, rotating, adjustbox, pict2e,array}
\usepackage{amsfonts}
\usepackage{longtable}
\usepackage{tabularx}
\usepackage{caption}
\usepackage{booktabs}
\usepackage{amsmath}
\usepackage{etoolbox}
\usepackage{makecell}
\usepackage{blindtext}
\usepackage[all]{xy}
\usepackage[utf8]{inputenc}
\usepackage[normalem]{ulem}
\makeatletter
\numberwithin{equation}{section}
\numberwithin{figure}{section}
\usetikzlibrary{arrows, positioning, decorations.pathmorphing, decorations.pathreplacing, decorations.markings, matrix, patterns, snakes}
\hypersetup{colorlinks=true}
\hypersetup{linkcolor=black}
\hypersetup{citecolor=black}
\hypersetup{urlcolor=black}
\theoremstyle{plain}
\newtheorem*{thm*}{\protect\theoremname}
\theoremstyle{definition}
\newtheorem*{defn*}{\protect\definitionname}
\providecommand{\definitionname}{Definition}
\providecommand{\theoremname}{Theorem}
\makeatother
\tikzset{
  big arrow/.style={
    decoration={markings,mark=at position 1 with {\arrow[scale=1.5,#1]{>}}},
    postaction={decorate},
    shorten >=0.4pt},
  big arrow/.default=black}

    {\endtcolorbox}
\usepackage{mdframed}
\mdfdefinestyle{statementstyle}{
   frametitlerule = true,
   frametitlefont = {\normalfont\bfseries},
   frametitlebackgroundcolor = blue!10,
}
\mdtheorem[style=statementstyle]{statement}{Stringy thermalization}
\begin{document}
\begin{titlepage}
\hfill \\
\vspace*{15mm}
\begin{center}
{\Large \bf High energy scattering and string/black hole transition}

\vspace*{15mm}

{\large Alek Bedroya\footnote{\href{mailto:abedroya@g.harvard.edu}{abedroya@g.harvard.edu}}}
\vspace*{8mm}

\parbox{ \linewidth}{\begin{center}Jefferson Physical Laboratory, Harvard University,\\ Cambridge, MA 02138, USA\end{center}}\\

\vspace*{0.7cm}

\end{center}
\begin{abstract}
We give a short review of high energy scattering in string theory, thermal properties of strings, explain the connection between the two and how they all tie together with the string/black hole transition.  
\end{abstract}

\end{titlepage}

\tableofcontents

\section{Introduction}

Most of our non-perturbative understanding of quantum gravity comes from dualities and black holes. What makes these two phenomena special is that they allows us to perform a perturbative calculation for strongly coupled objects of the theory. In this note we will focus specifically on black holes. In the case of black holes, Hawking's field theory calculation gives the density of states at deep UV where the states have strong gravitational screening (i.e. they are black holes). 

In the absence of a complete theory of quantum gravity, it is challenging to sharply define what "quantum gravity" means. However, we can expect a quantum theory gravity in Minkowski spacetime to provide a consistent description of particles\footnote{Here by particle we refer to any weakly coupled state that could be understood by a pole in the scattering amplitude.} and their high-energy scattering amplitudes, especially those of graviton. Any such description would be naturally limited since at some energy scale $\Lambda_{BH}$, the massive would-be-particles are replaced by black holes. String theory incorporates this transition in a very elegant way. But it is natural to ask, what do we expect from a "quantum gravity" beyond the energy scale $\Lambda_{BH}$ when black holes become relevant? The good news about black hole is that many of their properties follow from IR calculations and are universal.  We can use those properties as a litmus test for any theory of quantum gravity. This note focuses on how string theory passes black hole related litmus tests.

Before going further, let us clarify the difference between a particle and a black hole. For a particle, the effective field theory breaks before its would-be horizon. For example, take the electron. Due to the screening effect (or the running of the charge), the classical Maxwell equations are only reliable for length scales greater than its Compton wavelength, which is much greater than electron's Schwarzschild radius. Therefore, we call the electron a \textit{particle} rather than a black hole. Similarly, for strings, the effective field theory can only be trusted for length scales greater than the string length $l_s$, which is much greater than the Schwarzschild radius for masses $m\ll (l_sg_s^2)^{-1}$, where $g_s$ is the string coupling. In this sense, we can think of black holes as would-be particles that can be described by effective field theory all the way up to their horizons, and perhaps beyond that. This suggests that in a  theory of quantum gravity, we need to adopt a frame-work that includes both black holes and particles, where at some mass scale the two transition to each other. In string theory specifically, the light particles are described by excitations of a string. Therefore, we expect a transition between string states and black holes. We can test this continuity in multiple ways. One is to look for thermal properties of black holes in strings, and another is to look for scattering properties of strings in black holes.

The first approach is commonly referred to as string/black hole transition, however, the second approach is equally important given that scattering amplitudes is a big chunk of the data that constitutes what we call string theory. In this note we try to review some properties of strings and black holes both from thermodynamic as well as scattering theory. One of the main goals of this note is to present some of the existing literature in a coherent way that connects it to the string/black hole transition.

The organization of this paper is as follows. In section \ref{st} we review the common lore and some recent developments about string/black hole transition from the matching of their thermal properties. In section \ref{ss} we study the string/black hole transition in light of scattering amplitudes. We continue that study in more detail in section \ref{ss2} and discuss some of the nuanced differences between fixed-angle scattering and fixed impact parameter scattering that often leads to contradictory statements in the existing literature. We argue that the impact parameter is a more suitable approach to understand the string/black hole transition and opens new doors for further studies. In section \ref{CS}, we discuss some of the challenges for interpreting the string/black hole transition as a correspondence between back hole states and string states.

\section{String/black hole transition: entropy}\label{st}

Let us start with reviewing the string/black hole transition. The transition refers to the matching of multiple energy scales which suggests a phase transition between a weakly interacting strings and black holes \cite{Horowitz:1996nw,Susskind:1993ws} . 
\begin{itemize}
    \item The energy scale at which the Schwarzschild radius of the string is the string length. At this energy scale, the Einstein field equations are reliable at the Schwarzschild radius. Therefore, the string excitations, beyond this energy scale, if they exist, must be described by black holes. The Schwarzschild radius of a string excitation with mass $M$ is given by 
    \begin{align}
        r_S\sim (m l_s^{d-2}g_s^2)^{\frac{1}{d-3}}.
    \end{align}
    The crossover energy scale is where $r_S\sim l_s$ which corresponds to the mass
    \begin{align}
        M_c\sim M_s/g_s^2,
    \end{align}
    where $M_s=l_s^{-1}$ is the string mass.
    \item Energy at which the number of free string microstates of a given energy matches those of a black hole of the same energy. It was shown in the early years of string theory that the leading term in the entropy scales linearly with energy (see \cite{Huang:1970iq} and references therein). Here, we give a simple combinatorial argument for unfamiliar readers. The mass of a string excitation at high energies is proportional to $\sqrt{N}$ where $N=\sum_i i\cdot n_i$ is the level and $n_i$ is overall excitation number of all the $i$-th modes. Let us first find a lower bound for the number of microstates. Consider a string at level $N$, where the only exited modes are those with $i<\sqrt{N}$. Given that any single excitation of such modes can at most contribute $\sqrt{N}$ to the level, we need the sum of the excitation numbers $n_i$ to be at least $\sqrt{N}$. Consider an arbitrary set of $n_i\geq0$ with $1<i<\sqrt{N}$ that add up to $\simeq \sqrt{N}$, and choose $n_1$ to account for the rest of the level. 
    \begin{align}
        n_1\simeq N-\sum_{i>1}i\cdot n_i.
    \end{align}
    These states are only a subset of total string excitations, but they are very easy to count. Their number is equal to the number of different ways that one can partition $\simeq \sqrt{N}$ into $\simeq \sqrt{N}$ numbers which is given by $\sim {2\sqrt{N}\choose \sqrt{N}}>2^{2\sqrt{N}}/(2\sqrt{N})$. Therefore, the number of microstates are expected to increase exponentially with $M\sim M_s\sqrt{N}$. Upon doing a more careful calculation, one can see that the states that we did not consider ($i>\sqrt{N}$), do not alter the exponential behavior. This is because, although considering the modes $i>\sqrt{N}$ will increase the number of possibilities. Each such mode will have a very large contribution to the level and thus bring down the total number of remaining excitations. The leading contribution to the number of string states with mass $M$ goes like $\text{N}(M)\propto (M/M_s)^{-\gamma}\exp(\delta\cdot(M/M_s))$ for some positive constants $\gamma$ and $\delta$ \cite{Mertens:2015ola}.

    So far we calculated the number of single string microstate. A more accurate comparison would be to compute the number of microstate with energy $M$ in a free string gas. The number of microstates with $k$ strings each of energy $\sim E/k$ is $\simeq \text{N}(E/k)^k\sim (\frac{E}{M_s})^{-k\gamma}\exp(-\delta\cdot(E/M_s))$. Due to the polynomial prefactor, the number of single string states ($k=1$) dominates the microstates and the leading term in the entropy of the string gas is,
    \begin{align}
        S_{string}(M)\simeq \beta_H M,
    \end{align}
where $\beta_H\sim 1/M_s$ is a constant. It is easy to see that the free string partition function $\int dM S(M)e^{-\beta M}$ diverges at $\beta=\beta_H$ which is called the Hagedorn temperature. This divergence of the partition function can be used for a more precise evaluation of $\beta_H$ in different string theories \cite{Atick:1988si,OBrien:1987kzw,Kogan:1987jd,Sathiapalan:1986db}. 

Now let us calculate the entropy of a d-dimensional black hole in string units. The black hole entropy is proportional to its area in Planck units which goes like 
\begin{align}
    r_S^{D-2}\sim M^\frac{d-2}{d-3},
\end{align}
in Planck units. After going to the string units via $l_P^{d-2}=l_S^{d-2}g_s^2$, we find
    \begin{align}
        S_\text{black hole}\sim (g_s^2l_s^{d-2}M^{d-2})^{\frac{1}{d-3}}.
    \end{align}
    By equating the two entropies we find that the crossover energy scale is 
    \begin{align}
        M_c\sim M_s/g_s^2,
    \end{align}
    regardless of the spacetime dimension.
\end{itemize}

The matching of all these energy scales paints a consistent picture. It suggests that in the corners of moduli space where string theories are reliable, there is transition between strings and black hole. The transition happens at $M\sim M_s/g_s^2$. The particles with masses below this mass are interacting weakly enough to be described perturbatively by string theory. More importantly, their Schwarzschild radius is smaller than the string length and therefore it cannot be probed using scattering. On the other hand, "\textit{particles}" with masses higher than $M_s/g_s^2$ are no longer weakly interacting. They have strong gravitational screening to the point that they collapse into black holes. Moreover, the black hole geometry near the horizon is reliable for them given their large enough radii. 

In the above discussion we discussed the continuous transition in the entropy as we go from  free string gas to black holes. However, there are other macroscopic quantities that we can compare between the two and one of them is the size of the system. The size of a free string with mass $M$ is $l\sim l_s\sqrt{M/M_s}$. It is easy to see that this size does not match that of the black hole at the transition point. However, this is not an issue because, as we increase the energy of a string, we expect the free string approximation to break down. The self-gravitational interaction of the string should gradually set in to the point where after the transition, the correct description of the thermal system is in terms of black holes. 

As we approach the Hagedorn temperature $T_H$, the string gas will start backreacting on the background and curve it towards becoming a black hole. We can study this curved background by looking at string partition function. One way to understand a classical background of a black hole is that if we Wick rotate the background into Euclidean signature, it would determine the free energy of the corresponding thermal system. One can calculate the free energy of black hole as a thermal system using this trick. 

Going back to strings, to find the background sourced by a self-gravitating string, we can calculate its thermal partition function and see if corresponds to any Euclidean $d$ dimensional background. It was shown in \cite{Polchinski:1985zf} that the thermal partition of strings can be calculated by evaluating its vacuum amplitude on $\mathbb{R}^{d-1}\times S^1$ where the radius of $S^1$ at spatial infinity goes to $1/(2\pi T)$ where $T$ is the temperature of the black hole. We can estimate this partition function by looking at its saddle point. We can think of this saddle point as the configuration that solves the equations of motion when we compactify Euclidean $d$-dimensional string theory on a circle with anti-periodic boundary condition for fermions. String theory on a small circle is well-approximated by a $(d-1)$ dimensional field theory. Thus, in order to find the saddle point, we can extremize the $(d-1)$ dimensional effective action in the non-compact dimensions. Let us review the light fields that would admit a field theory description. The $(d-1)$ dimensional field theory would contain the light fields resulted from the usual dimensional reduction of higher dimensional fields. However, near the Hagedorn temperature, there will be a new light field. This is the string winding mode. At temperatures above the Hagedorn temperature, this mode becomes tachyonic which signals a breakdown in the stringy treatment and a phase transition. The resulting background was found by Horowitz and Polchinski for temperatures near the Hagedorn temperature in dimensions $d<7$ \cite{Horowitz:1997jc}. To connect this background to the black hole background, one can think of the $d-1$ non-compact dimensions as the analog of the $d-1$ non-compact dimensions of Euclidean black holes. However, here we have an additional light $(d-1)$ dimensional field which is the winding mode. Unlike the black hole which is localized at a point in the Euclidean signature, this solution is spread out and it is reasonable to expect that the Lorentzian continuation of this solution would have a size of the same order. It turns out in rotating the Euclidean background to a Lorentzian background, the field values exit the perturbative regime of string theory and is therefore, not trustable. But as far as thermodynamic properties are concerned, we can use the Euclidean solution to do calculations.

In \cite{Chen:2021dsw}, authors did precise calculations for the size and entropy of the Horowitz-Polchinski solution and demonstrated that it provides a continuous bridge between free strings and black holes. However, they found a surprising feature for the solution. In type II theories, it seems that there is no smooth transition between this background and the black hole background, if they are viewed as the target space of the string worldsheet CFT. However, this does not invalidate the string/black hole transition as it could be that along the transition, the purterbative description of the worldsheet breaks down.

\section{String/black hole transition: scattering amplitudes}\label{ss} 

Another way that we can study the string/black hole transition is through scattering amplitudes. One of the main features of black holes is their universal thermal properties. Since quantum gravity in Minkowski space is usually formulated in terms of an S-matrix, it is natural to expect that the thermal properties of black holes have some meaning in terms of the scattering amplitudes. We review an argument that postulates that at small impact parameters, high energy scattering events go through black hole formation/evaporation \cite{Banks:1999gd}. In that case, since black hole evaporate thermally, we expect the amplitude for a particular out-going state to be $\sim \exp(-cS_{BH})$. But this is only at small enough impact parameters and large energies that allow black hole formations. Fortunately, we can calculate high energy scattering amplitudes below that energy threshold with string theory. In that case, scattering processes go though string states rather than black hole states. If string black hole transition is correct, we expect a smooth transition between the amplitude. In fact, this transition has been demonstrated which is what we will review in this section. 

\subsection{Scattering through black holes} 

Consider a 2-particle in-going state with a center of mass energy $E$ and impact parameter $b$. Suppose $E$ is much larger than the mass of the smallest black hole, and $b$ is much smaller than the Schwarzschild radius of a black hole with mass $E$. In that case, when the particles distance from each other becomes much smaller than the Schwarzschild radius, we can still trust the effective field theory description which tells us that the geometry collapses into a black hole. 

Now the outcome of this scattering event is determined by the evaporation of the black hole. Due to the thermodynamic nature of black holes, we expect that the amplitude of a typical 2 particle out-going state to be $\sim\exp(-S_{BH})$. However, even though the Hawking radiation at finite time interval is thermal, assuming the black hole evaporation respects unitarity, the overall out-going state must be pure and cannot be thermal. In other words, it is only when we look at part of the Hawking radiation and trace over the rest of the quantum system  that we expect the radiation to be thermal. Therefore, the overall amplitude is not exactly given by a black body radiation ensemble. However, we can break up the Hawking radiation into smaller intervals (early radiation and late radiation) and approximate the density matrix of each with an almost thermal density matrix with entropy $\propto S_{BH}$. In that case, a $2\rightarrow 2$ corresponds to the event where each part of the radiation is a single particle. This allows us to estimate the amplitude of a $2\rightarrow 2$ event as 
\begin{align}
    \mathcal{M}(2\rightarrow 2)\sim \exp(-c(E/M_P)^\frac{d-2}{d-3}),
\end{align}
where $c\sim \mathcal{O}(\log E)$ accounts for any polynomial correction. The above argument is explained more precisely in Appendix \ref{A1}.

\subsection{Gross Mende saddle points}

In the previous subsection we saw that the thermodynamic properties of black holes gives a universal behavior for the scattering amplitudes at impact parameters $b\ll r_S(E)$ where $r_S$ is the Schwarzschild radius of a black hole with mass $E$. This analysis is only valid when the center of mass energy is larger than the mass of the smallest black hole. From our discussion of string/black hole transition, we expect the mass of the smallest black hole to be $\sim M_s/g_s^2$. Such a black hole would have a Schwarzschild radius of $l_s$ which is the smallest length scale that the perturbative string can detect. Therefore, the black hole analysis, in conjunction with the string/black hole transition suggests that the scattering amplitude of events with impact parameter $b\sim l_s$ and energies $E\sim M_s/g_s^2$ is given by 

\begin{align}\label{BHSA}
    \mathcal{M}\sim \exp(c(\frac{M_s/g_s^2}{M_P})^\frac{d-2}{d-3})\sim \exp(cg_s^{-2}).
\end{align}

For smaller energies, we can trust string perturbation and if the transition is correct, we should be able to reproduce the above scattering amplitude from string theory. In the following, we review the work of Gross and Mende which leads to the following equation for string amplitudes.
\begin{align}
    \mathcal{M}\sim  \exp(c(E/M_s)),
\end{align}
where $c\sim\mathcal{O}(\ln(E))$. As one can see, the above equation reproduces the black hole result of \eqref{BHSA} at energies $E\sim M_s/g_s^2$.

Gross and Mende solved for the saddle points of the string wordlsheet amplitudes \cite{Gross:1987ar,Gross:1987kza}. Once these saddle points are found, we can think of them as the classical string processes whose amplitudes receive quantum corrections from other worldsheet configurations.

Gross-Mende saddle points are K-sheet covers of the sphere with $N$ branch points that are connected by branch cuts. The branch-points correspond to the external particles. They showed that such worldsheet saddle points correspond to the solutions to 2d electrostatic on the sphere and the total energy corresponds to the worldsheet action. The multi-sheet covers correspond to fractional charges for which the electric potential will not be single valued. The genus of the corresponding worldsheet can be written in terms of $K$ and $N$ as follows. 
\begin{align}
    G=\frac{1}{2}(K-1)(N-2),
\end{align}
In the following we focus on the $2\rightarrow 2$ scattering for which $N=4$. Given that higher $K$ corresponds to smaller fractional electric charges, we expect the worldsheet action to be inversely proportional to $K$. Therefore, using the result for the tree-level string amplitude, we find that the amplitude of the Gross-Mende saddle point behaves like
\begin{align}
    \exp(-S_{K})\sim \exp(-\frac{1}{4K}(s\ln(s)+t\ln(t)+u\ln(u))),
\end{align}
in string units. The full string amplitude contains the sum over all of these classical saddles and goes as 
\begin{align}
   \mathcal{M}\sim \sum_{K} g_s^{2G-2}e^{\frac{1}{4K}(s\ln(s)+t\ln(t)+u\ln(u))+\mathcal{O}(\ln(E))}.
\end{align}
Mende and Ooguri showed that this series is Borel resummable \cite{Mende:1989wt}. However, without reserving to the full summation, we can easily estimate the order of magnitude of this series by looking at the term with the largest contribution. The largest contribution comes from $G\sim E=\sqrt{s}$ and the corresponding amplitude is 
\begin{align}\label{GMSR}
       \mathcal{M}\sim e^{-c(E/M_s)}.
\end{align}

The Gross-Mende saddle points correspond to a scattering processes where the intermediate state is $K$ coincident strings. The impact parameter of the scattering event can be estimated by the size of the intermediate strings which goes as $E/K$. For the scattering event with the dominant contribution we have $K\sim E/M_s$ which means the impact parameter is of the order of string length. 
\begin{align}
    b_{\text{GM}}\sim l_s.
\end{align}
Since this is the Schwarzschild radius of the smallest black hole, we expect the Gross-Mende result to match with the black hole result at the transition point. As one can see this is indeed the case. 

\begin{align}
    \mathcal{M}_{string}(2\rightarrow 2)&\sim \exp(-c(E/M_s)),\nonumber\\
    \mathcal{M}_{BH}(2\rightarrow 2)&\sim \exp(-c(E/M_P)^\frac{d-2}{d-3}),
\end{align}
and at the transition point $E\sim M_s/g_s^2$ we have
\begin{align}
    \ln(\mathcal{M}_{string})\sim g_s^{-2}\sim \ln ( \mathcal{M}_{BH}).
\end{align}
What is perhaps more surprising is that both results go like $\exp(-S_{ent})$ of the corresponding thermal system. We will come back to this point in the next subsection.

In \cite{Dvali:2014ila}, the above matching between the stringy calculation and the black hole result was generalized to $2\rightarrow N$ scattering.

\subsection{Relationship between scattering amplitudes and entropy} \label{RAC}
    
As we saw in the previous subsection, high energy scattering amplitude at small enough impact parameters always seem to go as $\exp(-cS)$ where $S$ is the thermodynamic entropy of the system at energy $E$. For energies below $M_s/g_s^2$, $S$ is the entropy of a string gas, while for energies $E>M_s/g_s^2$, it is the entropy of a black hole with mass $E$. 

In the black hole case, there is a clear interpretation for the relationship between the scattering amplitude and the entropy. We can think of black holes as highly complex quantum systems where the information of the incoming state gets scrambled in the black hole Hilbert space and creates a thermal ensemble. Let us call this feature of scattering amplitudes, \textit{thermalization} \footnote{Note that the outgoing state is still pure and not thermal. However, when restricted to small enough subsystems in the out-going Hilbert space, the density matrix looks thermal.}.

The fact that the scattering amplitudes go as $\exp(-cS)$ in the stringy regime suggest that some form of thermalization happens in the stringy regime as well. In the following, we will try to understand this stringy thermalization from high energy scattering of strings.

If thermalization as described above happens, we expect that the string scattering at low impact parameters ($b\sim l_s$) to mostly go through a localized stringy system with $S_{ent}$ degrees of freedoms. However, $S_{ent}$ is the entropy of the string gas and it is unclear why a string gas must form. Moreover, the microstates of a free string gas are dominated by the 1-particle string states, however, Gross-Mende analysis suggest that string amplitudes are dominated by saddle points with a large number of intermidiate strings. So how do we resolve this inconsistency and explain the appearance of the entropy in the scattering amplitude? The key feature is that even though the states with higher number of strings have polynomially less microstates, they can have an entropy of the same order. This is beacuse entropy is the logarithm of the number of microstates. So it could be that the entropy appearing in the scattering amplitude is not that of the full string gas, but a subsystem with a large number of strings with a similar entropy. For example, the entropy of a localized $k$-string gas with energy $E$ is \cite{Mertens:2015ola}
\begin{align}
    S_k(E)\simeq \ln[(\frac{E}{M_s})^{-d\cdot k}\exp(\beta_H E)],
\end{align}
where the equality is up to subleading corrections. for $k\lesssim E$, we have
\begin{align}
    S_k(E)\simeq S_{string}(E).
\end{align}
However, for larger values of $k$ the entropy quickly drops. Therefore, the string/black hole transition suggests the following.

\begin{statement*}Most of the string scattering events with impact parameters $b\sim l_s$ and energy $E\gg M_s$, go through classical worldsheet saddles with $\lesssim E$ number of strings. The entropy associated with Hilbert space of $k\lesssim E$ strings is of the same order as the free string gas entropy.
\end{statement*}

This notion of thermalization carries over to black hole's thermodynamic properties after the string/black hole transition. The worldsheet saddles are the stringy version of high energy scatterings with $E>M_s/g_s^2$ and $b<r_S$ that go through the classical saddle of black hole formation. Just as the scattering amplitudes in the black hole case is inversely proportional to the black hole microstates. The high energy string amplitude is inversely proportional to the microstates with $k\lesssim E$ strings.

Now let us check the above hypothesis for string thermalization using the Gross-Mende saddle points. In the Gross-Mende saddle points, the impact parameter goes as $E/K$. Therefore, the processes with impact parameter less than $l_s$ correspond to processes with less than $E$ strings that have an entropy comparable to the free-string gas. Therfore, the stringy amplitude is expected to exhibit thermal properties even pre-transition. To see this more precisely, the contribution of a Gross-Mende saddle is 
\begin{align}
   \sim g_s^{2G}\exp(-cE^2/k),
\end{align}
where $2G=(k-1)(N-2)$ is the genus and $N$ is the number of external particles. Among $k\lesssim E$ configurations which have impact parameters $b\sim l_s$, the amplitude is dominated by $K\sim E$ 
\begin{align}
    \mathcal{M}(2\rightarrow N)\sim \exp(-cE/M_s)
\end{align}
as we would expect from the entropy argument. Therefore, we see that for all the low impact parameter processes, the dominant contribution goes through a classical saddle with an entropy comparable to the thermodynamic entropy. This is to say that the intermediate states are starting to explore more and more of the Hilbert space as the complexity of the time-evolution is increasing. The increase in the complexity was shown to manifest itself in the form of erratic behavior of the string S-matrix for highly excited strings \cite{Gross:2021gsj}.

One can think of our argument above as an evidence for Hawking-like radiation for $k$ string states where the average energy of the strings is $E/k \gtrsim>M_s$. The reason we expect this thermal radiation, is that the entropy of this system is proportional to that of the string gas. So another test for the string/black hole transition would be to see if a single energetic string emits a black-body radiation. This was calculated and confirmed in \cite{Kawamoto:2013fza} which makes another strong case for the string/black hole transition.

\section{High energy scattering}\label{ss2} 

\subsection{fixed angle vs fixed impact parameter} 

So far, whenever we have talked about high energy scattering, it has been about scattering events with small impact parameters. In this subsection we discuss fixed angle scattering and how black holes manifest themselves in the fixed angle scattering. Let us consider a generic $2\rightarrow 2$ scattering scattering amplitude. In addition to the scattering energy $\sqrt{s}$, we can choose one of impact parameter $b$ or scattering angle $\theta\simeq 2\sin^{-1}(\sqrt{-t/s})$ to describe the out-going state, but not both. This is due to the fact that impact parameter is a measure of angular momentum, and it does not commute with the momenta of the out-going particles.

In the following, we often use the Mandelstam variable $t$ instead of the scattering angle $\theta\simeq 2\sin^{-1}(\sqrt{-t/s})$. The transformation between amplitudes in $(s,t)$ space and $(s,b)$ space is given by the following Fourier transform \cite{Amati:1988tn}.
\begin{align}\label{tb}
    \frac{1}{s}A(s,t)=(\epsilon_a\cdot\epsilon_d)(\epsilon_b\cdot\epsilon_c)\cdot4\int d^{D-2}b e^{i\bf{q\cdot b}}a(s,b),
\end{align}
where $\bf{q}$ is the momentum exchange $t=-\bf{q}^2$. In the above formula, $\epsilon_i$ are the polarization vectors of the external particles and $a(s,b)$ is the polarization-independent part of the amplitude. As one can see, a fixed angle scattering involves a complex averaging over the fixed impact parameter amplitudes.

For large impact parameters, we can think of impact parameter as a measure of angular momentum $l\simeq Eb/2$. However, for small impact parameters $b\ll E^{-1}$, the quantization of the angular momentum becomes important and this approximation breaks down. The partial wave expansion of the fixed angle amplitude is the analog of equation \eqref{tb} for $l$. The $d$-dimensional partial wave expansion is given by \cite{Soldate:1986mk}
\begin{align}\label{gfa}
    \mathcal{A}(s,t)=2^{2d-3}\pi^\frac{d-3}{2}\Gamma(\frac{d-3}{2})s^\frac{4-d}{2}\sum_{l=0}^\infty (l+\frac{d-3}{2})C_l^\frac{d-3}{2}(\cos(\theta))f_l(s),
\end{align}
where $C_l^\frac{d-3}{2}$ are Gegenbauer polynomials. 

For energies larger than the black hole threshold $E\gg M_s/g_s^2$, and impact parameters smaller than the Schwarschild radius $b\simeq 2l/E\ll r_S(E)$, we expect the cross-section to be exponentially small due to black hole formation and evaporation. 
\begin{align}
    b\simeq 2l/E\ll r_S(E):~~\ln(a(s,b))\sim \ln(f_l(E))\sim -S_{BH}(E).
\end{align}
However, if the amplitude is polynomial for impact parameters greater than the Schwarzschild radius, the overall sum \eqref{tb} or \eqref{gfa} comes out polynomial. This is in fact what happens. For large enough impact parameters one can use the Born approximation to find a polynomial amplitude. Therefore, the overall fixed angle amplitude is expected to be polynomial. Another way to see this polynomial behavior is that due to the exponential suppression of the amplitude $f_l(E)$ for $b<r_S(E)$, the partial waves can no longer interfere to create a delta function. Now the overall amplitude looks like that of a scattering off of a black disc with radius $\sim r_S(E)$. The shadow effect from the edge of the disc gives a polynomial cross-section for any given angle. This effect was calculated more precisely in \cite{Giddings:2009gj}.

Note that there are multiple different regimes between the black hole regime and the Born regime based on the value of $\ln(b)/\ln(E)$. We will discuss these different regimes in more detail in the next subsection.

The above discussion shows that the fixed angle scattering amplitude is much less sensitive to the black hole physics than the fixed impact parameter amplitudes. In the next subsection we will focus on finding the perturbative scattering amplitudes in the E--b plane.

Let us take a moment to comment on the Gross-Mende saddles. In principle, since the Gross-Mende solution is a saddle, we can use it regardless of what parameters we use for the scattering. However, as we will see in the next subsection, when the scattering angle is too small, there is another contribution from higher impact parameters that has a more dominant contribution. Therefore, the Gross-Mende result should be only used to estimate the fixed angle scattering at large enough angles. There is a very intuitive explanation for this. As we discussed earlier, the Gross-Mende saddle with dominant contribution has impact parameter $\lesssim l_s$ which suggests that it is the stringy version of the black hole formation/evaporation. Scattering events that go through black hole formation/evaporation are hard scattering events that emit particles at all angles. However, scatterings at larger impact parameters are expected to be softer and have a more focused amplitudes at small scattering angles. Therefore, just like the black hole example, we should expect the small angle scattering to always be dominated by soft processes that do not go through any saddle (Gross-Mende or black hole). 

In the previous section, we tested the string/black hole transition by studying the continuity of the amplitude across the $E\sim M_s/g_s^2,~b\sim l_s$ point. However, when we look at scattering amplitudes in terms of the $E-b$ plane, we see that the black hole region has another boundary at $ E>M_s/g_s^2,~~b\sim r_S(E)$. 

\begin{figure}[H]
    \centering

\tikzset{every picture/.style={line width=0.75pt}} 

\begin{tikzpicture}[x=0.75pt,y=0.75pt,yscale=-1,xscale=1]

\draw  (103,378.99) -- (536.14,378.99)(146.31,22.01) -- (146.31,418.65) (529.14,373.99) -- (536.14,378.99) -- (529.14,383.99) (141.31,29.01) -- (146.31,22.01) -- (151.31,29.01)  ;
\draw    (310.06,211.15) -- (310.77,379.17) ;
\draw  [dash pattern={on 4.5pt off 4.5pt}]  (146.67,211.61) -- (310.06,211.15) ;
\draw    (482.61,17.42) -- (310.06,211.15) ;

\draw (402.14,393.87) node [anchor=north west][inner sep=0.75pt]    {$r_S( E) \sim \left( Eg_{S}^{2} l_{S}^{D-2}\right)^{1/[ D-3]}$};
\draw (105.3,18.04) node [anchor=north west][inner sep=0.75pt]    {$b$};
\draw (416.94,63.32) node [anchor=north west][inner sep=0.75pt]  [rotate=-311]  {$b\sim R( E)$};
\draw (124.32,202.51) node [anchor=north west][inner sep=0.75pt]    {$l_{s}$};
\draw (305.32,392.57) node [anchor=north west][inner sep=0.75pt]    {$l_{s}$};
\draw (386.78,267.7) node [anchor=north west][inner sep=0.75pt]    {$a( s,b) \sim \exp( -cS_{BH})$};
\draw (189.33,182.31) node [anchor=north west][inner sep=0.75pt]   [align=left] {\begin{minipage}[lt]{57.72pt}\setlength\topsep{0pt}
\begin{center}
Perturbative\\regime
\end{center}

\end{minipage}};
\draw (399.55,184.8) node [anchor=north west][inner sep=0.75pt]   [align=left] {\begin{minipage}[lt]{78.69pt}\setlength\topsep{0pt}
\begin{center}
Non-perturbative\\regime
\end{center}

\end{minipage}};
\draw (292.76,331.74) node [anchor=north west][inner sep=0.75pt]  [rotate=-270] [align=left] {Boundary I};
\draw (330.72,158.29) node [anchor=north west][inner sep=0.75pt]  [rotate=-311] [align=left] {Boundary II};

\end{tikzpicture}
    \caption{The scattering in the non-perturbative regime is dominated by black hole formation/evaporation which leads to exponentially suppressed amplitudes. The scattering amplitude in the perturbative regime can be calculated using string theory. One can check the matching of the stringy result and the black hole result across the boundary I as a non-trivial test of string/black hole transition.}
\end{figure}
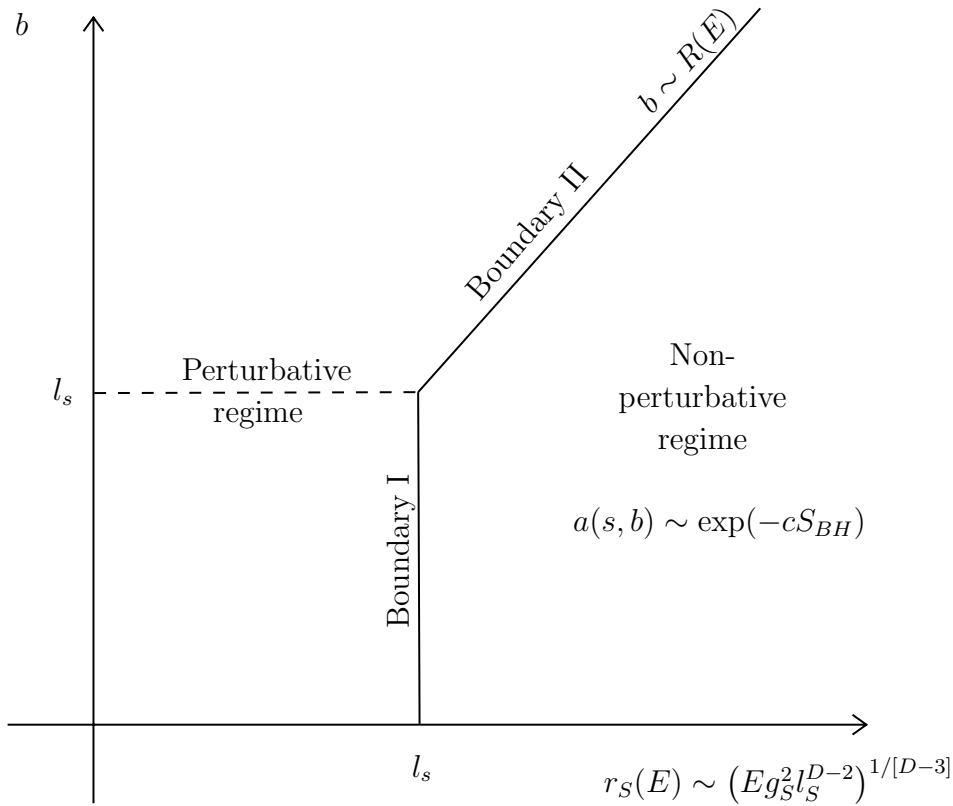

One could hope that the scattering amplitudes at larger impact parameters are purterbutively calculable and one can check the string/black hole via continuity of the scattering amplitudes across this boundary as well. In a series of papers \cite{Amati:1987uf,Amati:1988tn,Fabbrichesi:1993kz,Veneziano:2004er,Amati:2007ak,Veneziano:2008zb,Veneziano:2008xa,Addazi:2016ksu}, Veneziano et al. developed a method to reformulate string perturbation theory in terms of $E$ and $b$ and calculate the string amplitude in the $b\gg r_S(E)$ regime. We will review these results in the next subsection and discuss their connection with the string/black hole transition. 

\subsection{ACV approach}

First let us summarize the Amati-Ciafaloni-Veneziano (ACV) approach. The idea is to first calculate the string theory amplitudes in the Regge limit $t$-fixed and $s\rightarrow\infty$, and express those results in terms of the impact parameter. Then we try to push the approximation toward the fixed angle scattering regime as much as possible.

ACV show that similar to the Eikonal regime in field theory where the ladder diagrams become dominant, in string theory the diagrams factorize as well. In particular, the genus $h$ amplitude goes like

\begin{align}
    A^{(h)}(s,t)\rightarrow \epsilon_a\cdot\epsilon_d\epsilon_b\cdot\epsilon_c4 (\frac{i}{2s})^h &\int\frac{d^{D-2}\bf{q}_1...d^{D-2}\bf{q}_{h+1}}{(h+1)!(2\pi)^{(D-2)h}}\delta^{D-2}(\bf{q}-\sum_i\bf{q}_i)\nonumber\\
    &\times \Pi_{j=1}^{j=h+1}a_{tree}(s,t_j)\langle 0|\int\Pi_j\frac{d\sigma_j^ud\sigma_j^d}{(2\pi)^2}:e^{i\bf{q}_j\cdot(\hat X^u(\sigma_j^u)-\hat X^d(\sigma_j^d)}:|0\rangle.
\end{align}
Note that the operator $:e^{i\bf{q}_j\cdot(\hat X^u(\sigma_j^u)-\hat X^d(\sigma_j^d)}:$ captures the effect of stringy excitations that are absent in the field theory calculations. If we sum over $h$ in the above formula and plug in the output in \eqref{tb} we find a genus expansion for $a(s,b)$,
\begin{align}
    a^{(h)}(s,b)=\frac{(2i)^h}{(h+1)!}\langle 0|\hat\delta^{h+1}|0\rangle,
\end{align}
where,
\begin{align}
    \hat\delta&=\int\frac{d\bf{q}}{(2\pi)^{D-2})}\frac{a_{tree}(s,t)}{s}\int\langle 0|\int\frac{d\sigma_ud\sigma_d}{(2\pi)^2}:e^{i\bf{q}\cdot(\hat X^u(\sigma_u)-\hat X^d(\sigma_d)}:|0\rangle.\nonumber\\
    &=\int\frac{d\sigma_ud\sigma_d}{(2\pi)^2}:a_{tree}(s,b+\hat X^u(\sigma_u)-\hat X^d(\sigma_d)):.
\end{align}
Therefore, by summing over genus $h$ we find
\begin{align}
    a(s,b)=\sum_{h\geq 0} a^{(h)}(s,b)=\langle 0|\frac{1}{2i}(e^{2i\hat\delta}-1)|0\rangle.
\end{align}

We need to compute the vacuum expectation value of $e^{2i\hat\delta}$. For that we can start with by approximating that expectation value with $\hat X=0$. This approximation ignores string excitations and therefore must reproduce the result of graviton ladder diagrams in field theory calculation, i.e., the normal gravitational deflection.

We can read $\delta(b,s)$ from the tree-level amplitude (Veneziano amplitude) as follows.
\begin{align}
    \delta(b,s)\equiv \hat \delta|_{\hat X=0}=\frac{\pi g^2s}{(4\pi)^{D/2}}\int^{\infty+i\epsilon}_{-\infty-i\epsilon}\frac{dy}{y^{D/2-1}}\exp(-\frac{b^2}{4y})\times F(s/2,s/2;1;\exp(-y)).
\end{align}
The above integral has very different behaviors depending on $b/\sqrt{Y}$ where $Y=\ln(s)$\footnote{Remarkably, if one keeps track of the logarithmic dependence of c in \eqref{GMSR}, the same $\sqrt{Y}$ quantity shows up in the analysis of the Gross-Mende saddle points}. For $b\gg \sqrt{Y}$ we find,
\begin{align}
    \text{Re} \delta&\simeq \frac{g^2s}{8\pi}\frac{1}{\Omega_{D-4}}b^{D-4}\nonumber\\
    \text{Im} \delta&\simeq \frac{\pi g^2 s}{8(4\pi Y)^{D/2-2}}\exp(\frac{b^2}{4Y}).
\end{align}
From this point on we refer to the above expressions as elastic expressions since it ignores string excitations. For $b\ll \sqrt{Y}$ we find,
\begin{align}\label{inel}
    \text{Re} \delta&\simeq \frac{g^2s}{8\pi}\frac{1}{(4\pi Y)^{D/2-2}}(\frac{1}{D-4}-\frac{b^2}{4Y(D-2)}+...)\nonumber\\
    \text{Im} \delta&\simeq \frac{\pi g^2 s}{8(4\pi Y)^{D/2-2}}\exp(\frac{b^2}{4Y}).
\end{align}
We can enhance our approximation by looking at the quantum fluctuations of $\hat\delta$. 
\begin{align}
\exp(i\delta)\exp(i\Delta_\bot: \bar{\hat {X}}_{\bot u^2}:i\Delta_\parallel: \bar{\hat {X}}_{\parallel u^2}:+(u\leftrightarrow d)),
\end{align}
where $\Delta_{ij}=\partial^2 \delta/\partial b_i\partial b_j$.

\begin{align}
    \Delta_\parallel=-(D-3)\Delta_\bot=\frac{(D-3)g^2s}{4\Omega_{D-2}b^{D-2}}.
\end{align}

Using our results for elastic scattering we find for $b\gg\sqrt{Y}$
\begin{align}
    \text{Im}\delta=\text{Im}\delta_{el}+\frac{(D-3)\pi g^2s}{4\Omega_{D-2}b^{D-2}}.
\end{align}
The transition between the dominance of $\text{Im}\delta_{el}$ and $\frac{(D-3)\pi g^2s}{4\Omega_{D-2}b^{D-2}}$ happens at
\begin{align}
    b_{DE}=(\frac{2\pi^2(D-3)}{\Omega_{D-2}}sg^2)^\frac{1}{D-2}l_s^{\frac{D}{D-2}}.
\end{align}
Before calculating the fixed angle scattering amplitudes, let us summarize the above results in the $(s,b)$ plane so far. For $b\gg b_{DE}$, $a(s,b)$ is polynomial while for $\sqrt{Y}\ll b<b_{DE}$, we have $a(s,b)\sim \exp(- Bg^2s/b^{D-2})$. Finally, for $b\ll\sqrt{\ln(s)}$, from \eqref{inel}, we have $a(s,b)\sim \exp(-A g^2s/\ln(s)^{D/2-2})$.

\begin{figure}[H]
    \centering

\tikzset{every picture/.style={line width=0.75pt}} 

\begin{tikzpicture}[x=0.75pt,y=0.75pt,yscale=-1,xscale=1]

\draw  (12,408.8) -- (627,408.8)(73.5,20) -- (73.5,452) (620,403.8) -- (627,408.8) -- (620,413.8) (68.5,27) -- (73.5,20) -- (78.5,27)  ;
\draw    (306,226) -- (307,409) ;
\draw    (118,225) -- (306,226) ;
\draw    (551,15) -- (306,226) ;
\draw    (175,5) .. controls (145,66) and (121,156) .. (118,225) ;
\draw    (74,225) -- (118,225) ;

\draw (472,426.4) node [anchor=north west][inner sep=0.75pt]    {$R( E) \sim \left( Eg_{S}^{2} l_{S}^{D-2}\right)^{1/[ D-3]}$};
\draw (18,16.4) node [anchor=north west][inner sep=0.75pt]    {$b$};
\draw (119,296.4) node [anchor=north west][inner sep=0.75pt]  [font=\small]  {$A\sim \exp\left( -\ \frac{Ag_{s}^{2} s}{\ln( s)^{\frac{D}{2} -1}} \ \right)$};
\draw (223,61.4) node [anchor=north west][inner sep=0.75pt]    {$A\sim \exp\left( -\ \frac{Ag_{s}^{2} s}{b^{D-2}} \ \right)$};
\draw (82,29.4) node [anchor=north west][inner sep=0.75pt]    {$A\sim \mathcal{O}( 1)$};
\draw (466.57,58.68) node [anchor=north west][inner sep=0.75pt]  [rotate=-320.13]  {$b\sim R( E)$};
\draw (45,217.4) node [anchor=north west][inner sep=0.75pt]    {$l_{s}$};
\draw (302,424.4) node [anchor=north west][inner sep=0.75pt]    {$l_{s}$};
\draw (95.61,179.88) node [anchor=north west][inner sep=0.75pt]  [rotate=-282.71]  {$b^{D-2} \sim E^{2} g_{s}^{2} l{_{s}}^{D}$};
\draw (418,306.4) node [anchor=north west][inner sep=0.75pt]    {$A\sim \exp( -S_{ent})$};

\end{tikzpicture}
\end{figure}

In order to calculate the fixed angle amplitude \eqref{tb} we can use saddle point approximation. So for every angle $\theta$, first we find the impact parameter $\bf{b_s}$ where
\begin{align}
    q_i=-\frac{\text{Re} \delta}{\partial b_i}|_{\bf {b=b_s}} ~~~,~~~ \theta=\frac{|\bf{q}|}{\sqrt{s}}=-\frac{2}{\sqrt{s}}\frac{\text{Re}\delta}{\partial b}|_{\bf{b=b_s}}.
\end{align}
Then the fixed angle amplitude is given by 
\begin{align}
    A(s,t)\sim \exp(i\bf{q.b_s})e^{i\delta(b_s,E)},
\end{align}
up to polynomial prefactors. It turns out such a saddle point only exists for $\theta\sim l_s^{D-2}Eg_s^2/b^{D-3}$. In the $\sqrt{Y}\ll b<b_{DE}$ regime we have 
\begin{align}
    \theta=\frac{D-2}{D-3}(\frac{r_S(E)}{b})^{D-3},
\end{align}
where $r_S(E)=(Eg_s^2l_s^{D-2})^{1/(D-3)}$ is the corresponding Schwarzschild radius.

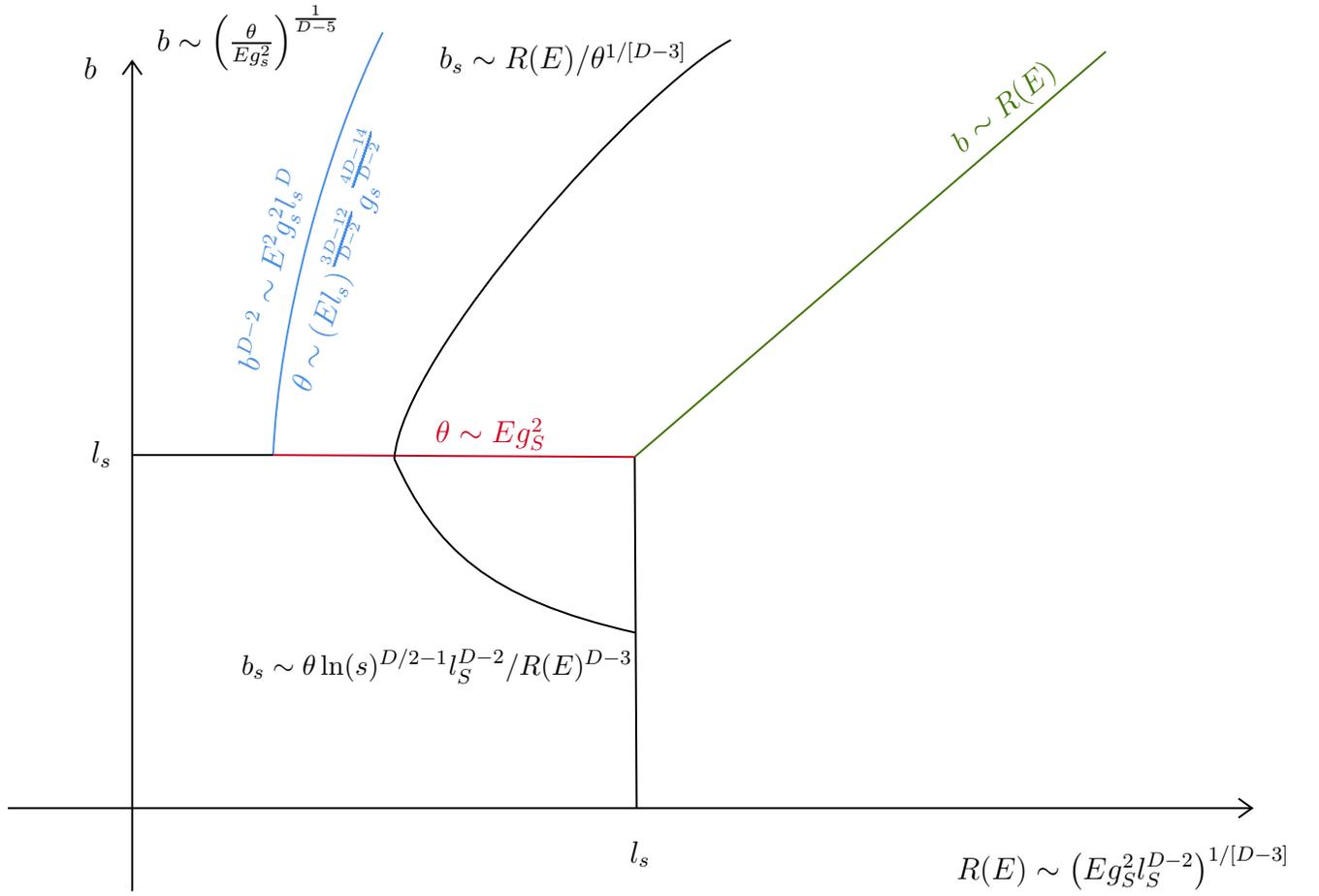
\begin{figure}[H]
    \centering

\tikzset{every picture/.style={line width=0.75pt}} 

\begin{tikzpicture}[x=0.75pt,y=0.75pt,yscale=-1,xscale=1]

\draw  (-6,429.8) -- (641,429.8)(58.7,41) -- (58.7,473) (634,424.8) -- (641,429.8) -- (634,434.8) (53.7,48) -- (58.7,41) -- (63.7,48)  ;
\draw    (320,247) -- (321,430) ;
\draw [color={rgb, 255:red, 208; green, 2; blue, 27 }  ,draw opacity=1 ]   (132,246) -- (320,247) ;
\draw [color={rgb, 255:red, 65; green, 117; blue, 5 }  ,draw opacity=1 ]   (565,36) -- (320,247) ;
\draw [color={rgb, 255:red, 74; green, 144; blue, 226 }  ,draw opacity=1 ]   (189,26) .. controls (159,87) and (135,177) .. (132,246) ;
\draw    (59,246) -- (132,246) ;
\draw    (195,248) .. controls (199,193) and (324,54) .. (370,30) ;
\draw    (195,248) .. controls (216,294) and (243,321) .. (320.5,338.5) ;

\draw (486,447.4) node [anchor=north west][inner sep=0.75pt]    {$R( E) \sim \left( Eg_{S}^{2} l_{S}^{D-2}\right)^{1/[ D-3]}$};
\draw (32,37.4) node [anchor=north west][inner sep=0.75pt]    {$b$};
\draw (114,345.4) node [anchor=north west][inner sep=0.75pt]  [font=\small]  {$b_{s} \sim \theta \ln( s)^{D/2-1} l_{S}^{D-2} /R( E)^{D-3}$};
\draw (480.57,79.68) node [anchor=north west][inner sep=0.75pt]  [color={rgb, 255:red, 65; green, 117; blue, 5 }  ,opacity=1 ,rotate=-320.13]  {$b\sim R( E)$};
\draw (36,237.4) node [anchor=north west][inner sep=0.75pt]    {$l_{s}$};
\draw (316,445.4) node [anchor=north west][inner sep=0.75pt]    {$l_{s}$};
\draw (109.61,200.88) node [anchor=north west][inner sep=0.75pt]  [color={rgb, 255:red, 74; green, 144; blue, 226 }  ,opacity=1 ,rotate=-282.71]  {$b^{D-2} \sim E^{2} g_{s}^{2} l{_{s}}^{D}$};
\draw (217,29.4) node [anchor=north west][inner sep=0.75pt]    {$b_{s} \sim R( E) /\theta ^{1/[ D-3]}$};
\draw (215,225.4) node [anchor=north west][inner sep=0.75pt]  [color={rgb, 255:red, 208; green, 2; blue, 27 }  ,opacity=1 ]  {$\theta \sim Eg_{S}^{2}$};
\draw (129.61,207.88) node [anchor=north west][inner sep=0.75pt]  [color={rgb, 255:red, 74; green, 144; blue, 226 }  ,opacity=1 ,rotate=-288.54]  {$\theta \sim ( El_{s})^{\frac{3D-12}{D-2}} g_{s}^{\frac{4D-14}{D-2}}$};
\draw (70,10.4) node [anchor=north west][inner sep=0.75pt]    {$b\sim \left(\frac{\theta }{Eg_{s}^{2}}\right)^{\frac{1}{D-5}}$};

\end{tikzpicture}
\end{figure}

The amplitude would then read 
\begin{align}
    A(s,\theta)\sim \exp(-Cg_s^\frac{-2}{D-3}(E/M_s)^\frac{D-4}{D-3}\theta^{-\frac{D-2}{D-3}}).
\end{align}
Although there is also a saddle point in $b<\sqrt{Y}$, 
\begin{align}
    \theta\sim\frac{bEg_s^2}{Y^{D/2-1}}
\end{align}
its contribution to the fixed angle amplitude is not dominant since the amplitude goes like $\sim \exp(-A g^2s/\ln(s)^{D/2-2})$ which is very similar to the Gross-Mende saddle point. However, Gross-Mende saddle point exists for all angles as opposed to ACV saddle points. It is important to note that the dominant Gross-Mende saddle point is also at $b<\sqrt{Y}$. 

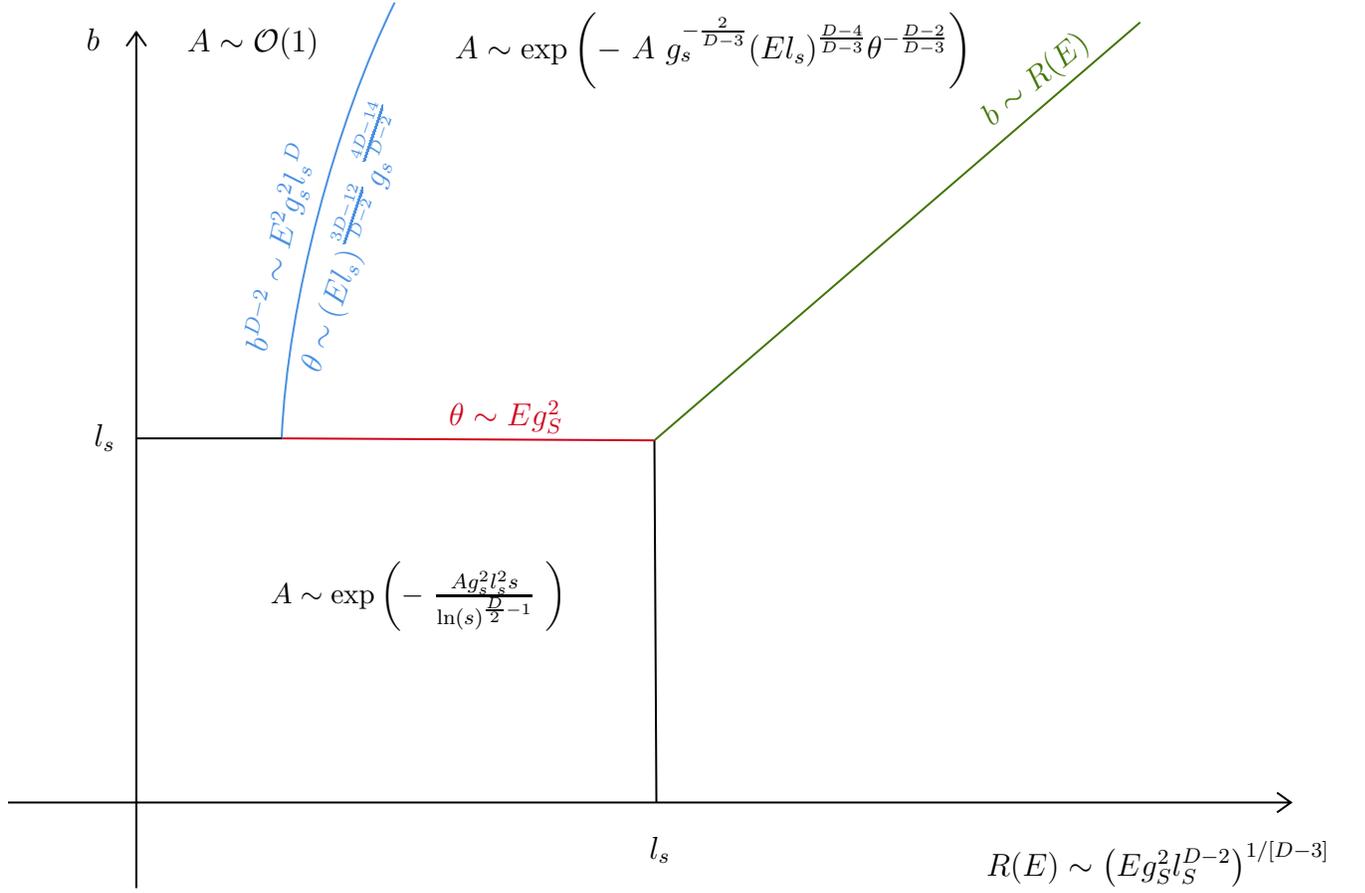
\begin{figure}[H]
    \centering

\tikzset{every picture/.style={line width=0.75pt}} 

\begin{tikzpicture}[x=0.75pt,y=0.75pt,yscale=-1,xscale=1]

\draw  (-6,429.8) -- (641,429.8)(58.7,41) -- (58.7,473) (634,424.8) -- (641,429.8) -- (634,434.8) (53.7,48) -- (58.7,41) -- (63.7,48)  ;
\draw    (320,247) -- (321,430) ;
\draw [color={rgb, 255:red, 208; green, 2; blue, 27 }  ,draw opacity=1 ]   (132,246) -- (320,247) ;
\draw [color={rgb, 255:red, 65; green, 117; blue, 5 }  ,draw opacity=1 ]   (565,36) -- (320,247) ;
\draw [color={rgb, 255:red, 74; green, 144; blue, 226 }  ,draw opacity=1 ]   (189,26) .. controls (159,87) and (135,177) .. (132,246) ;
\draw    (59,246) -- (132,246) ;

\draw (486,447.4) node [anchor=north west][inner sep=0.75pt]    {$R( E) \sim \left( Eg_{S}^{2} l_{S}^{D-2}\right)^{1/[ D-3]}$};
\draw (32,37.4) node [anchor=north west][inner sep=0.75pt]    {$b$};
\draw (480.57,79.68) node [anchor=north west][inner sep=0.75pt]  [color={rgb, 255:red, 65; green, 117; blue, 5 }  ,opacity=1 ,rotate=-320.13]  {$b\sim R( E)$};
\draw (36,237.4) node [anchor=north west][inner sep=0.75pt]    {$l_{s}$};
\draw (316,445.4) node [anchor=north west][inner sep=0.75pt]    {$l_{s}$};
\draw (109.61,200.88) node [anchor=north west][inner sep=0.75pt]  [color={rgb, 255:red, 74; green, 144; blue, 226 }  ,opacity=1 ,rotate=-282.71]  {$b^{D-2} \sim E^{2} g_{s}^{2} l{_{s}}^{D}$};
\draw (215,225.4) node [anchor=north west][inner sep=0.75pt]  [color={rgb, 255:red, 208; green, 2; blue, 27 }  ,opacity=1 ]  {$\theta \sim Eg_{S}^{2}$};
\draw (129.61,207.88) node [anchor=north west][inner sep=0.75pt]  [color={rgb, 255:red, 74; green, 144; blue, 226 }  ,opacity=1 ,rotate=-288.54]  {$\theta \sim ( El_{s})^{\frac{3D-12}{D-2}} g_{s}^{\frac{4D-14}{D-2}}$};
\draw (125,306.4) node [anchor=north west][inner sep=0.75pt]  [font=\small]  {$A\sim \exp\left( -\ \frac{Ag_{s}^{2} l_{s}^{2} s}{\ln( s)^{\frac{D}{2} -1}} \ \right)$};
\draw (83,37.4) node [anchor=north west][inner sep=0.75pt]    {$A\sim \mathcal{O}( 1)$};
\draw (218,29.4) node [anchor=north west][inner sep=0.75pt]    {$A\sim \exp\left( -\ A\ g_{s}^{-\frac{2}{D-3}}( El_{s})^{\frac{D-4}{D-3}} \theta ^{-\frac{D-2}{D-3}}\right)$};

\end{tikzpicture}
\end{figure}

According to the ACV analysis, Gross-Mende result is a saddle point that contributes to the fixed angle scattering but is not the dominant contribution. It most naturally compares to the ACV saddles found for $b<\sqrt{Y}$ where the amplitude exponentially decays. In this picture the amplitude for small impact parameters goes like $\exp (-S_{ent})$ and changes continiously accross the $E=M_s/g_s^2$ line.

Note that depending on the dimension, there could also be a minimum on the scattering angle of the saddle point in the region of exponential decay. When 
\begin{align}
    \theta\lesssim \theta_{DE}=E^{\frac{3D-12}{D-2}}g_s^{\frac{4D-14}{D-2}},
\end{align}
the saddle point lies in the elastic regime where amplitude is polynomial. In 4d, the LHS is $\theta_{DE}=g_s$ while in higher dimensions it grows with energy. Thus, in 4d, the ACV analysis suggests exponentially decaying amplitudes for fixed angle $\theta>g_s$ scattering.

\begin{align}
    A\sim \exp(-CE^\frac{D-4}{D-3}).
\end{align}
However, in higher dimesnions, for sufficiently high energies the amplitude is always polynomial.

The reason behind this mismatch in four dimensions is that ACV by its use of the saddle point approximation focuses on the phase of the amplitude rather than the absolute value of it for different values of $b$. In other words, ACV calculation assumes that because of the fastly oscilating phase of amplitudes as a function of $b$ or $l$ in \eqref{tb} or \eqref{gfa}, the amplitudes cancel each other out at values of $b$ or $l$ far away from the saddle point.

\section{Transition or correspondence?}\label{CS}

In the previous sections, we studied two aspects of strings and black holes that continuously match across the transition point. Namely, we looked at the transition between the scattering amplitudes and entropies. However, there are two other tests of continuity which challenge the correspondence between strings and black holes in its naive form. Those are the continuity of size and greybody factors. In fact these two properties are very closely connected for any gravitational bound state. The challenges we discuss in this section do not question the string/black hole transition, whether thermodynamically or as dominant scattering resonances. However, they question a particular interpretation of that transition which identifies black hole microstates with string states at the transition point. We refer to this interpretation as string/black hole correspondence to differentiate it from the more conservative observation of string/black hole transition.

\subsection{Size}

The size of a free string of energy $E$ which is given by $\sqrt{\langle{\overline{(X-X_{\text{avg}})^2}}\rangle}$ can be approximated as $l\propto l_s\sqrt{E\cdot l_s}$ \cite{Mitchell:1987hr,Mitchell:1987th}. Therefore, at the transition scale $E\sim M_s/g_s^2$, the size of a free string is $l\sim l_s/g_s$. This length is much greater than the size of the corresponding black hole  which is $l_s$. There are two ways to resolve this mismatch.
\begin{enumerate}
    \item As we increase the mass of the string and approach the transition point, the self-gravitating effects become important and shrink the size of the string to match it with that of the black hole. 
    \item The states that continuously connect to black hole microstates at $M\sim M_s/g_s^2$ are non-perturbative objects rather than strings. However, because they have smaller number of microstates than strings for $M<M_s/g_s^2$, they do not significantly contribute to entropy or amplitudes below the transition point. In this picture, the string black hole transition is a transition between string microstates to other strongly coupled objects that become more thermodynamically likely at the transition point.
\end{enumerate}

\subsection{Greybody factor}

Highly massive strings and black holes both emit a near black-body radiation (for thermal radiation of heavy strings, see \cite{Kawamoto:2013fza}). However, there is a fundamental difference between the two which lies in the deviation from the black-body spectrum. Black hole's Hawking radiation comes with a grebody factor which suppresses the radiation of higher spin particles. The origin of the grey-body factor is the gravitational potential created around the black hole which partially blocks higher spin particles from escaping to infinity. Since a free string does not have such a structure around it, it lacks the greybody factors. For a detailed account of the mismatch between the emission rates of strings and black holes see \cite{Mathur:1997wb}. In fact, if the back reaction of the string creates a structure with a size that matches that of the black hole, the matching of the grey-body factors will automatically follow. This is because the greybody factors of a Schwarzschild black hole, are universal consequences of its spherically symmetric gravitational profile \cite{Das:1996we}. Therefore, the question of continuity of the size and the grey-body factors are in fact the same question.

In order to see how string theory can account for the greybody factor, it is useful to look at setups where the black holes microstates are known to correspond to BPS branes with some compact dimensions. In those setups, the lower-spin particles correspond to polarizations of higher-dimensional fields along the compact dimension. In other words, the existence of the compact dimension is responsible makes a fundamental distinction between particles of different spins (see \cite{Das:1996wn,Maldacena:1996ix} for more detail). However, such a structure is absent for a free string with no compact dimensions. 

In fact, as we saw in the previous section, in scatterings with energies below $M_s/g_s^2$, a full thermalization does not happen. In other words, the majority of the scattering amplitude comes from intermediate states with large number of strings, which make up a negligible part of the string gas Hilbert space. At least from the perspective of scattering amplitudes, it seems that the string gas Hilbert space does not continuously connect to black hole Hilbert space. 

\subsection{Horowitz-Polchinski solution and its limitations}

The Horowitz-Polchinski solution, which is often said to describe a self-gravitating string, is a Euclidean EFT background that bridges black holes and free strings in terms of size \cite{Horowitz:1997jc}. However, it is important to note a few points about it.
\begin{itemize}
    \item The Horowitz-Polchinski solution only exists in $d\leq6$ spacetime dimensions. Recently, it was argued that using a dimension expansion in $d-7$, one can extend the results to higher dimensions  \cite{Balthazar:2022hno}. However, the nature of the solution still remains different in dimensions smaller and larger than $7$.
    
    \item The transition between the Horowitz-Polchinski background and the black hole background as worldsheet theories, seems to be obstructed for type II theories \cite{Chen:2021dsw}. 
    
    \item The Horowitz-Polchinski solution can be thought of as a string propagating in a thermal background, rather than a single self-gravitating string. In other words, it does not necessarily tell us that the background itself is a single string. For example, the insertion of a D-brane in a background is captured by the worldsheet theory of the fundamental string via boundary conditions. However, the fact that the background is described by a worldsheet theory does not imply that the background is a single string. This is why we have been cautious to use the the phrase string/black hole transition rather than string/black hole correspondence, since the transition of parameters do not imply that we can view the black hole microstates as extension of string states. In fact, in the known examples where the microscopic origin of the BPS black holes are known, the microstates correspond to non-perturbative objects rather than strings \cite{Strominger:1996sh,Maldacena:1997de,Callan:1996dv,}. However, nonetheless, those non-perturbative objects can be understood as worldsheet theories in weak coupling limits. 
\end{itemize}

\section{Conclusions} 

We reviewed the string/black hole transition and how it can be understood in terms of the properties of high energy scattering amplitudes in string theory. These two topics are closely related, mainly due to the fact that all of the known formulations of quantum gravity are in terms of some boundary observables. In flat spacetime, these boundary observables take the form of an S-matrix with a rich structure. One example of an S-matrix that leads to a consistent theory of quantum gravity is string perturbation theory. Therefore, one should expect the universal properties of black holes to be manifested in string scattering amplitudes. 

For that, it is helpful to see black holes as resonances that dominate scattering amplitudes when the impact parameter is much less than the Schwarzschild radius. In this regime, the scattering is given by black hole formation and evaporation. Since the Hawking radiation is thermal at short time scales, the amplitude of a given event is exponentially suppressed as $\ln(\mathcal{M})\sim -S_{BH}$ where $S_{BH}$ is the black hole entropy. 

Even though string perturbation breaks down in the regime where black hole processes are dominant, we can hope to access its boundary from string perturbation and see how the black hole behavior starts to form at the boundary. The string/black hole transition suggests that the mass of the smallest black hole is $M_s/g_s^2$. Therefore, the black hole regime is $E\gg M_s/g_s^2$ and $b\ll r_s(E)$. This region has two boundaries, one at $E=M_s/g_s^2$ where string states supposedly start to form black holes, and the other is $b\sim r_s$ for $E\gg M_s/g_s^2$ where the black hole already has a sharp meaning in the effective field theory. As we cross the second boundary, we should expect a sharp change in the scattering amplitudes as they lose their angular dependence and become exponentially suppressed to match the Hawking radiation. This is indeed the case and the stringy results at $b\gg r_S$ does not match the black hole result at $b\ll r_S$. This makes it challenging to study large black holes from string theory amplitudes. 

However, for the $E\sim M_s/g_s^2$ boundary, we expect to see a continuous transition motivated by the first order phase transition between strings and black holes. This is indeed the case.

In the stringy regime ($E\lesssim M_s/g_s^2$), the scattering events are mediated by resonances that are described by classical Gross-Mende saddle points. These scattering events are mediated by $\sim E/M_s$ localized strings and impact parameter $\sim l_s$. This is very similar to the black hole regime ($E\gtrsim M_s/g_s^2$) where the scattering events are mediated by black holes that can be seen as the saddle point of the classical theory. Interestingly, the stringy result is very similar to the black hole result in that it is also given by $\mathcal{M}\sim\exp(-cS)$ where this time $S$ is the entropy of a string gas. A priori, the relationship between scattering amplitude and entropy is not clear in the stringy regime, because as opposed to black holes, the intermediate state does not thermalize along the scattering event. The explanation of this matching is a little subtle in the stringy regime. As we showed in \ref{RAC}, the entropy of the string gas with fixed $k\lesssim E/M_s$ number of strings, is of the same order of a string gas with energy $E$ which matches with the black hole entropy at the transition point. Based on this observation, we propose a thermal interpretation for the string/black hole transition. As we increase the energy, most of the scattering goes through saddles that correspond to a subsystem of entropy $\sim S$. Moreover, given that the dependence of the amplitude to the individual outgoing momenta is small, these are hard scattering processes that seem insensitive to the information of a individual out-going particles. In other words, the information is highly scrambled and is hidden in the entanglement between the particles. As we increase the energy, the Hilbert space that is accessible to the in-going states eventually is described by black holes. This fits nicely with the string/black hole transition and allows one to study the information paradox just below the transition using string theory S-matrix \cite{Veneziano:2004er}. Moreover, the string S-matrix opens the door to study the complexity of the black hole via the emergence of chaos in the high energy string scatterings \cite{Gross:2021gsj}. 

We also reviewed two different interpretations of the string/black hole transition in section \ref{CS}. The first interpretation views the string/black hole transition as a phase transition between string states and black hole states in terms of dominance in thermodynamics and scattering analysis. The second interpretation views the transition as an identification between the string states and the black hole microstates at the correspondence point. The first interpretation is more conservative and is supported by the known examples in string theory. This is due to the fact that the known microscopic descriptions of black hole microstates in string theory are not in terms of excitations of weakly strings, but non-perturbative objects such as D-branes. 

\section*{Acknowledgement}
We are grateful to Cumrun Vafa and Dieter Lüst for invaluable discussions that lead to this note. We also thank Samir Mathur for many useful discussions on the topic and valuable comments on the draft. This work is supported by a grant from the Simons Foundation (602883, CV) and by the NSF grant PHY-2013858.

\appendix

\section{Scattering amplitude and Hawking radiation}

In a unitary theory, the entropy of the black hole right after the collapse is zero. However, as the black hole evaporates, the entropy of the thermal radiation outside $S_R$ increases. For the overall state to remain pure, the Von Neumann entropy of the black hole $S_{BH}$ must increase accordingly. However, this entropy is different from the thermodynamic entropy of the black hole $S_{area}$, which is the maximal entropy of a black hole of a given mass and is given by the area law. 
\begin{align}
    S_R=S_{BH}\leq S_{area},
\end{align}
where the first equality is imposed by unitarity. 

Suppose we call the radiation coming out of the black hole before the Page transition the early time radiation and the rest of it the late time radiation. Before the Page transition, the LHS of the above inequality increases while the RHS decreases. However, at the Page time, the LHS catches up, and the inequality is saturated afterward. Therefore, after the Page transition, the black hole is described by a thermal density matrix since it has saturated its thermodynamic entropy.

The early time radiation and the late time radiation could be approximately separated by either their energies or their positions. For a black hole of mass $M$, the wave packets radiated before the Page time typically have energies $E\lesssim M$ at $r=\infty$ while the late time radiations have energies $E\gtrsim M$. This is because the temperature of the black hole grows much higher than the initial temperature after the Page time. However, perhaps a more natural separation is based on the position. The wavepackets radiated before the Page time have widths of $\sim r_S$. Therefore, a null wall with a spatial thickness of $\sim r_S$ can separate the early time radiation from the late time radiation. This allows us to decompose the radiation Hilbert space into a tensor product of the early time Hilbert space and a late time Hilbert space to a good precision. 

Now, let us talk about the density matrices of early time and late time radiations. The common lore about the Page curve gives us a good approximation for $\rho_{early}$ and $\rho_{late}$ individually. They each have an entropy of $\sim S_{area}(M)/2$. However, the entire state is not easy to find. For starters, the unitarity implies the overall state is pure and therefore is not $\rho_{early}\otimes\rho_{late}$. The early time and the late time radiations are individually thermal, however, when considered together, they are highly entangled, and the entanglement contains the information of the initial pure state. 

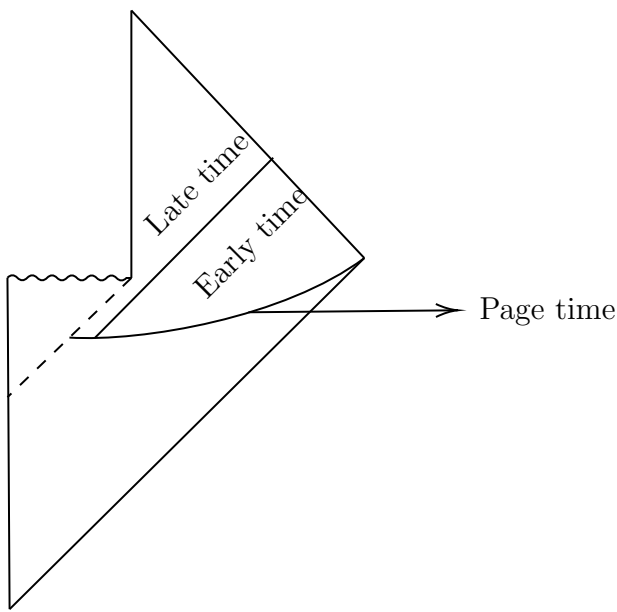
\begin{figure}
    \centering

\tikzset{every picture/.style={line width=0.75pt}} 

\begin{tikzpicture}[x=0.75pt,y=0.75pt,yscale=-1,xscale=1]

\draw    (219.47,151.57) -- (219.7,179.86) -- (220.55,318.86) ;
\draw    (219.47,151.57) .. controls (221.14,149.91) and (222.81,149.91) .. (224.47,151.58) .. controls (226.14,153.25) and (227.8,153.25) .. (229.47,151.59) .. controls (231.14,149.93) and (232.8,149.93) .. (234.47,151.6) .. controls (236.14,153.27) and (237.8,153.27) .. (239.47,151.61) .. controls (241.14,149.95) and (242.8,149.95) .. (244.47,151.62) .. controls (246.14,153.29) and (247.8,153.29) .. (249.47,151.63) .. controls (251.14,149.97) and (252.8,149.97) .. (254.47,151.64) .. controls (256.14,153.31) and (257.8,153.31) .. (259.47,151.65) .. controls (261.14,149.99) and (262.8,149.99) .. (264.47,151.66) .. controls (266.14,153.33) and (267.8,153.33) .. (269.47,151.67) .. controls (271.14,150.01) and (272.8,150.01) .. (274.47,151.68) .. controls (276.14,153.35) and (277.8,153.35) .. (279.47,151.69) -- (281.98,151.7) -- (281.98,151.7) ;
\draw    (281.97,16.65) -- (281.97,151.68) ;
\draw    (399.46,141.63) -- (220.55,318.86) ;
\draw    (281.98,16.67) -- (399.46,141.63) ;
\draw  [dash pattern={on 4.5pt off 4.5pt}]  (281.97,151.68) -- (219.49,211.86) ;
\draw    (353.5,91) -- (263.5,182) ;
\draw    (250.73,181.77) .. controls (287.5,184) and (359.46,171.63) .. (399.46,141.63) ;
\draw    (341,169) -- (444.5,168.02) ;
\draw [shift={(446.5,168)}, rotate = 179.46] [color={rgb, 255:red, 0; green, 0; blue, 0 }  ][line width=0.75]    (10.93,-3.29) .. controls (6.95,-1.4) and (3.31,-0.3) .. (0,0) .. controls (3.31,0.3) and (6.95,1.4) .. (10.93,3.29)   ;

\draw (310.09,154.88) node [anchor=north west][inner sep=0.75pt]  [rotate=-315] [align=left] {Early time};
\draw (285.36,123.12) node [anchor=north west][inner sep=0.75pt]  [rotate=-315] [align=left] {Late time};
\draw (456,160) node [anchor=north west][inner sep=0.75pt]   [align=left] {Page time};

\end{tikzpicture}
\caption{Early time and late time radiations separated by a null wall with spatial thickness $\sim r_S$ at spatial infinity.}
\end{figure}
\vspace{10pt} 

\textbf{\large Early time radiation:}

In the early time, the entropy of the radiation is much smaller than the dimension of the black hole. Therefore, Page's calculation shows that for a generic total pure state, the radiation coming out at any time interval is almost entirely un-entangled with any other time before the Page time. Therefore, a simple black body radiation calculation will lead to a good approximation for the density matrix.

\begin{align}\label{early}
    \rho_{early}\simeq\otimes_{t<t_{Page}}\rho_{black-body}[t,t+\delta t],
\end{align}
where $\rho_{black-body}[t,t+\delta t]$ is the density matrix of the black body radiation emitted over the time interval $[t,t+\delta t]$ where $M^3\gg\delta t\gg M$. The significance of this time interval is that in a time interval $\sim M$ an $\mathcal{O}(1)$ number particle of energy $\sim k_BT\sim 1/M$ are expected to be emitted. In other words, we are approximating the evaporation of the black hole's mass over time interval $\gg 1/M$ with a continuous decay of mass. This approximation clearly does not account for highly unlikely events where the early time radiation contains just a few particles with energies $\mathcal{O}(M)\gg\delta t$. This is something we will come back to later. However, note that since such events are very unlikely, the approximation \eqref{early} is a great approximation for the partition function and any thermodynamic quantity that can be derived from it.

\vspace{10pt} 

\textbf{\large Late time radiation:}

After the Page transition, the entropy of the black hole catches up with its Thermodynamic entropy. When that happens, the black hole is described by a thermal ensemble. We can leverage this fact to learn a lot about the density matrix of the late time radiation. 

During every short time interval $[t,t+\delta (t)]$ where $M(t)\ll\delta t\ll M^3$, a wave packet of black body radiation leaves the black hole, which leads to a slight decrease in the black hole's mass. However, if the black hole follows the Page curve, it remains at thermal equilibrium. This means the evaporation is adiabatic in the sense that it does not push the black hole out of equilibrium. 

The black body radiation formula tells us how to calculate the density matrix of the radiation coming out at any time interval $[t,\delta t]$. However, the important question is, can we use a similar formula to \eqref{early} for the overall density matrix of the late time radiation? The answer to this question is yes!

Take the first time interval after the Page transition. Since the relaxation time (or the scrambling time) is much smaller than the evaporation time, we can say the black hole density matrix $\rho_{BH}$ remains almost the same during the relaxation time. Open systems where this condition is satisfied, we can use Born approximation which says the state always remains separable $\rho_{BH}(t)\otimes\rho_{late}(t)$. Therefore, we can say that after every small period of time we have 
\begin{align}
    \rho_{BH}(t)\otimes \rho_{late}(t)\rightarrow\rho_{BH}(t+\Delta t)\otimes \rho_{Black body}[t,\Delta t]\otimes\rho'_{late}(t)
\end{align}
where $\rho'_{late}(t)$ represents the unitary time evolution of $\rho_{late}(t)$ with the same Von Nuemann entropy. Now, by induction we can conclude

\begin{align}\label{late}
    \rho_{late}\simeq\otimes_{t>t_{Page}}\rho_{black-body}[t,t+\delta t]
\end{align}

Thus, we know the distribution of the particles after and before the Page transition with very good accuracy. However, we do not know the joint probability distributions. For example, we know the early time radiation is $\ket\psi_1$ with probability $p_1$ and the state $\ket\psi_2$ describes the late time radiation with probability $p_2$. However, this does not mean that the probability of $\ket\psi_1\otimes\ket\psi_2$ as the total Hawking radiation is $p_1\times p_2$. This is because the overall radiation is not described by the tensor product $\rho_{early}\otimes\rho_{late}$. Rather, it is a pure state, and the choice of the pure state determines the correlation between early time and late time measurements. 

However, not all hope is lost! We can still say some useful things for an ensemble average of black hole microstates. For example, suppose we average overall in-going pure states that collapse into a black hole, then the probability would be $p_1\times p_2$. An easy way to see this is that such an ensemble should have a Von Neumann entropy of $S(M)$, which should remain unchanged under the unitary time evolution. However, this is the same entropy as of $\rho_{early}\otimes\rho_{late}$ since each of $\rho_{early}$ and $\rho_{late}$ have entropy of $S(M)/2$. 
\vspace{10pt}

\textbf{Calculation of the scattering amplitudes} \label{A1}
\vspace{10pt}

In this section, we use the details of the previous section to estimate the scattering cross-section at small impact parameters and large energies. 

Following we only focus on 2-2 scattering amplitude. We will combine the following results to bound the scattering cross-section in the strong-gravity regime. Suppose an intermediate black hole is formed from a collision of two states with energies $M/2$. Using the discussion of the last section, we estimate the probability of a black hole of mass $M$ to decay into the same in-going particles with matching momenta. Note that this event corresponds to the very unlikely event where we have a single particle of energy $M/2$ in the late time radiation. This means the radiation before the black hole reaches mass $M/2$ consists of a single particle, and the radiation after also consists of a single particle. Here, the natural dividing point in the evaporation process is the time when half of the black hole mass is evaporated. This time differs from the Page time by an $\mathcal{O}(1)$ factor. However, up to an $\mathcal{O}(1)$ factor, we can estimate the amplitude by computing the probability of the entirety of the early (or late time) radiation consisting of a single particle of energy $\mathcal{O}(M)$. Let us start with the early-time radiation. Suppose the partition function of the early time radiation is given by $\mathcal{Z}$. The probability of having a single particle of energy $M/2$ with specified momentum in the early time radiation is 

\begin{align}\label{FS}
    \Pr_{Early}\sim e^{-\beta \frac{M}{2}}/\mathcal{Z}_{early}.
\end{align}
We can use the equation \eqref{early} to estimate the above probability. Suppose $T(t)$ and $\mathcal{Z}(t)$ respectively represent the temperature and the partition function of $\rho_{Black body}[t,t+\delta t]$. Then, from the equation \eqref{early} we find
\begin{align}
    S\simeq\sum_t \frac{d}{dT(t)}(T(t)\ln \mathcal{Z}(t)).
\end{align}
All the temperatures in the early time radiation are of the order $\sim M^{-1/(d-3)}$ in Planck units where $d$ is the dimension of spacetime. Given that multiplying the temperature by a constant does not affect the individual terms, we can multiply all of the temperatures by $\mathcal{O}(1)$ constants to unify the derivatives as 
\begin{align}
        S\simeq\sum_t \frac{\partial}{\partial M^{-1/(d-3)}}(M^{-1/(d-3)}\ln \mathcal{Z}(M,t/M^3)).
\end{align}
Now by moving the sum inside the derivative, we find
\begin{align}
        S_{early}\simeq \frac{\partial}{\partial M^{-1/(d-3)}}(M^{-1/(d-3)}\ln \mathcal{Z}_{early}(M)).
\end{align}
After plugging in the entropy $S\propto M^\frac{d-2}{d-3}$ in the above equation we find
\begin{align}
    (d-3)AM^\frac{d-2}{d-3}\gtrsim\frac{d}{d M^{-1/(d-3)}}(M^{-1/(d-3)}\ln\mathcal{Z}_{early}),
\end{align}
where 
\begin{align}\label{Ac}
    A=\frac{2\pi}{(d-2)(d-3)}(\frac{8\Gamma(\frac{d-1}{2})}{(d-2)\pi^\frac{d-3}{2}})^\frac{1}{d-3}.
\end{align}
Integrating the above inequality leads to 
\begin{align}
    AM^\frac{d-2}{d-3}\sim\ln(\mathcal{Z}_{early}).
\end{align}
Plugging in $\mathcal{Z}_{early}$ from the above equation into \eqref{FS} leads to
\begin{align}
        \Pr_{early}\sim e^{-\frac{(d-2)}{2}AM^\frac{d-2}{d-3}- AM^\frac{d-2}{d-3}}\sim e^{\frac{d}{2}AM^\frac{d-2}{d-3}}\sim e^{\frac{d}{2}As^\frac{d-2}{2(d-3)}}
\end{align}
Note that for a generic pure state, we can say that the probability of the 2-2 scattering is less than the above bound 
\begin{align}
        \mathcal{M}(2\rightarrow 2)\lesssim e^{-\frac{d}{2}As^\frac{d-2}{2(d-3)}}
\end{align}

However, for the average ensemble where $\rho_{net}=\rho_{early}\otimes\rho_{late}$ we can find a stronger constraint because we expect to have two exponential factor like that of above which multiply to get
\begin{align}
        \mathcal{M}(2\rightarrow 2)\sim  e^{-dAs^\frac{d-2}{2(d-3)}}.
\end{align}
But this does not change the linear the dependence of the exponent in black hole entropy. Therefore, generically, we expect the scattering amplitude of $2\rightarrow 2$ events at energies $E$ much higher than the mass of the smallest black hole, and impact parameters smaller than the radius of a black hole with mass $E$ to go like 
\begin{align}
    \mathcal{M}\sim \exp(-cE^\frac{d-2}{d-3}),
\end{align}
where $c\lesssim\mathcal{O}(\log(E))$.
\bibliographystyle{unsrt}
\bibliography{References}
\end{document}